\documentclass[manuscript,screen]{acmart}
\usepackage{subcaption}

\AtBeginDocument{%
  }

\setcopyright{acmlicensed}
\copyrightyear{2023}
\acmYear{2023}
\acmDOI{XXXXXXX.XXXXXXX}

\acmConference[RecSys '23]{17th ACM Conference on Recommender Systems}{September 18--22,
  2023}{Singapore}
\acmBooktitle{17th ACM Conference on Recommender Systems (RecSys '23), September 18--22, 2023, Singapore}
\acmPrice{15.00}
\acmISBN{978-1-4503-XXXX-X/18/06}

\begin{document}

\title[Understanding User Behavior in Carousel Recommendation Systems]{Understanding User Behavior in Carousel Recommendation Systems for Click Modeling and Learning to Rank}
%

\author{Santiago de Leon-Martinez}
\affiliation{%
  \institution{Faculty of Information Technology, Brno University of Technology}
  \city{Brno}
  \country{Czechia}
}
\additionalaffiliation{%
  \institution{Kempelen Institute of Intelligent Technologies}
  \city{Bratislava}
  \country{Slovakia}
}
\email{santiago.deleon@kinit.sk}
\orcid{0000-0002-2109-9420}

\renewcommand{\shortauthors}{de Leon-Martinez}\textbf{}

\begin{abstract}
Carousels (also-known as multilists) have become the standard user interface for e-commerce platforms replacing the ranked list, the previous standard for recommender systems. While the research community has begun to focus on carousels, there are many unanswered questions and undeveloped areas when compared to the literature for ranked lists, which includes information retrieval research on the presentation of web search results. This work is an extended abstract for the RecSys 2023 Doctoral Symposium outlining a PhD project, with the main contribution of addressing the undeveloped areas in carousel recommenders: 1) the formulation of new click models and 2) learning to rank with click data. 

We present two significant barriers for this contribution and the field: lack of public datasets and lack of eye tracking user studies of browsing behavior. Clicks, the standard feedback collected by recommender systems, are insufficient to understand the whole interaction process of a user with a recommender requiring system designers to make assumptions, especially on browsing behavior. Eye tracking provides a means to elucidate the process and test these assumptions. Thus, to address these barriers and encourage future work, we will conduct an eye tracking user study within a carousel movie recommendation setting and make the dataset publicly available. Moreover, the insights learned on browsing behavior will help motivate the formulation of new click models and learning to rank. 
\end{abstract}



\maketitle

\section{Introduction}
\begin{figure*}[!tp]
  \begin{subfigure}[b]{0.6\columnwidth}
    \centering
    \includegraphics[width=\textwidth]{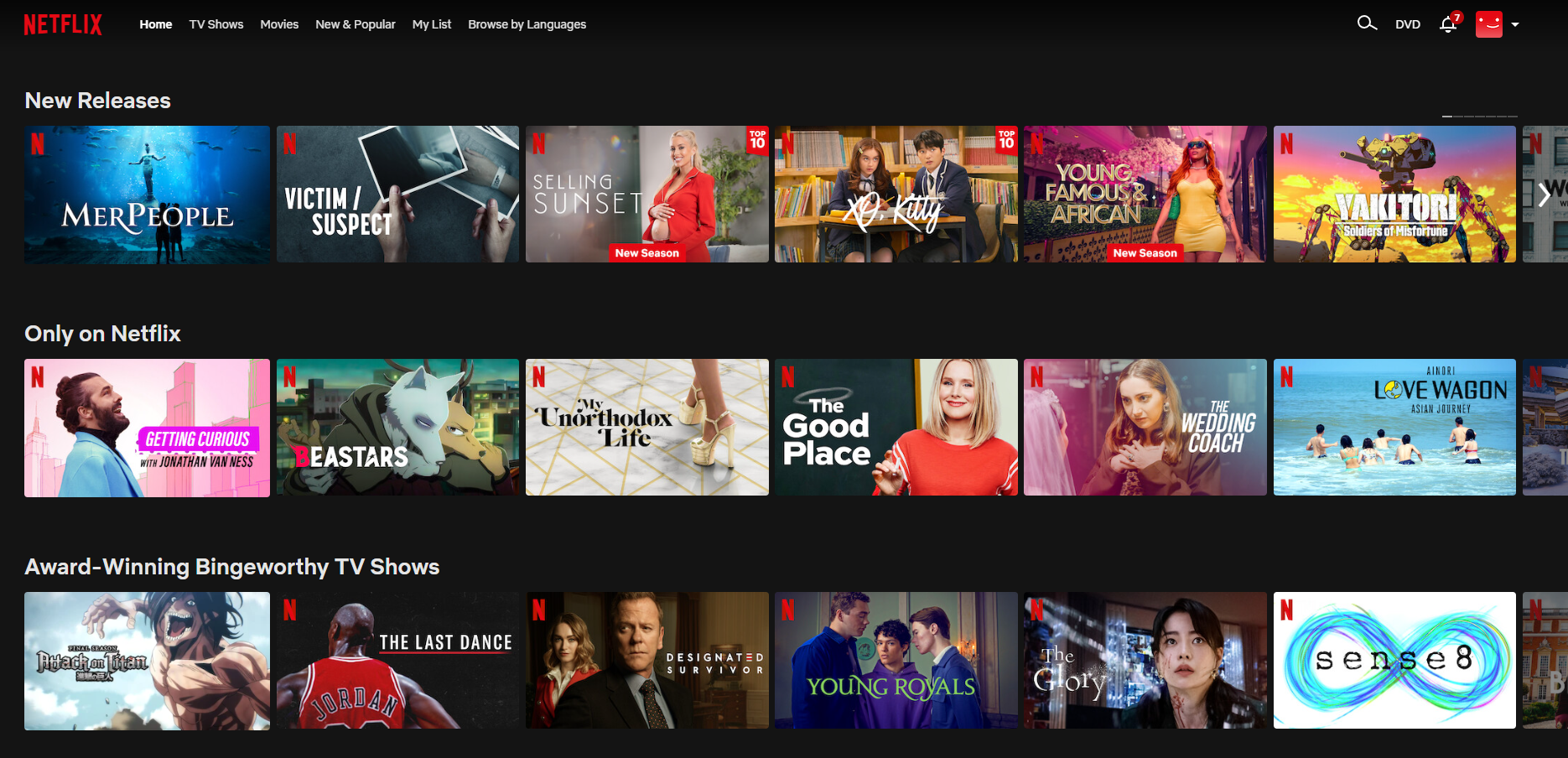}
    \label{fig:f1}
  \end{subfigure}
  \begin{subfigure}[b]{0.39\columnwidth}
    \centering
    \includegraphics[width=0.45\textwidth]{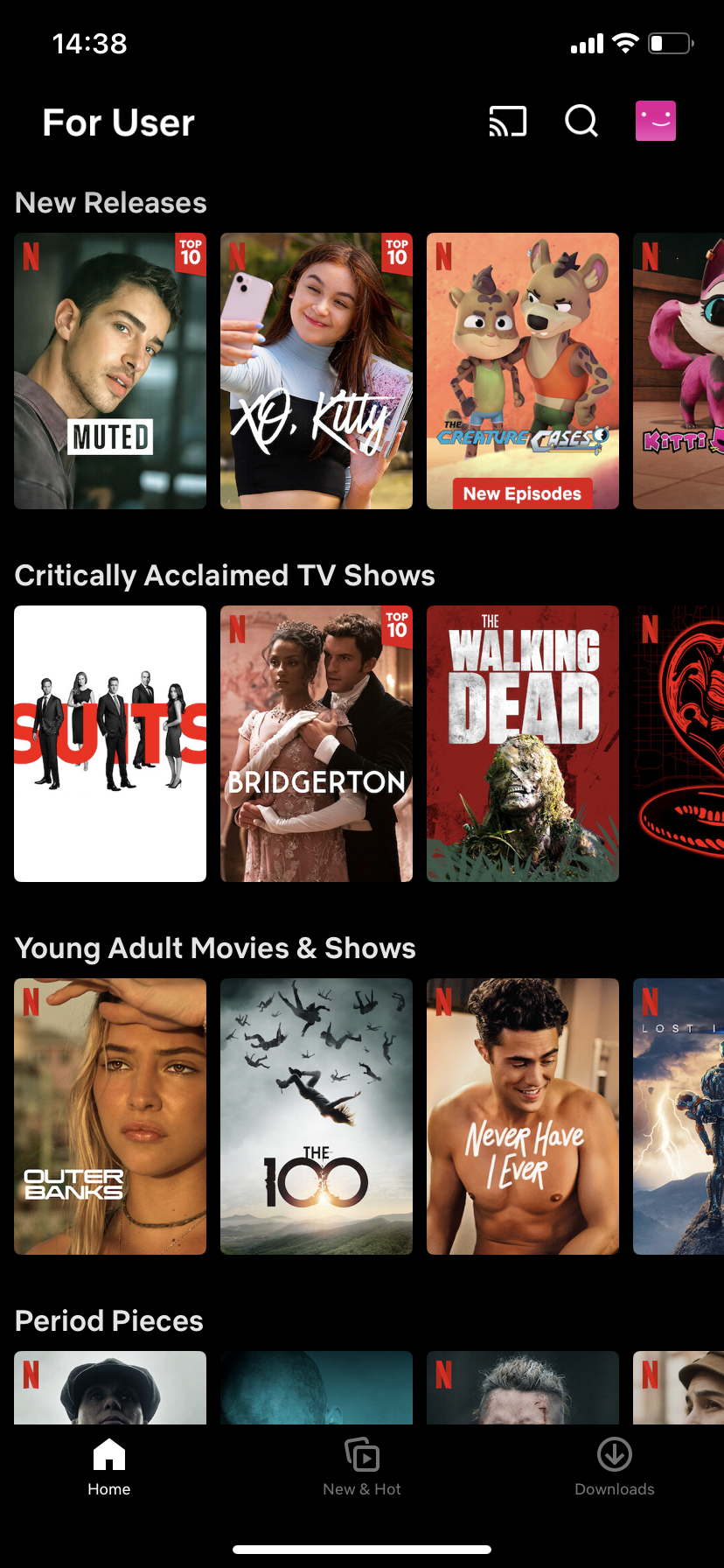}
    \label{fig:f2}
  \end{subfigure}
  \caption{Carousel interface used by Netflix homepage for desktop (left) and mobile (right)}
  \label{fig:big1}
\end{figure*}

Over the years, work in recommender systems has extensively explored and developed the single ranked list. However, more recently, \emph{carousels} (also-known as multilists) have become the most prominent user interfaces for recommendation platforms, especially in e-commerce, to the point where it is difficult to find non-carousel interfaces for marketplaces or streaming services today. This is true across mobile and desktop interfaces, such as Netflix seen in \textbf{Figure \ref{fig:big1}}, and other popular platforms, such as Amazon, Ebay, HBO, and Spotify, that all use carousels on their homepages and during exploration of predefined categories or subcategories. It is this area of initial impressions and exploration, where carousels are predominant, while specific searches still default to the single ranked list, whether in 1D top-down or 2D grid format. 

The term \textit{carousels} is used broadly in the literature (see e.g. \cite{rahdari_ranked_2022,rahdari_simulation-based_2022,jannach_exploring_2021, bendada_carousel_2020}). For clarity, we define carousels as multiple distinct lists with the following properties: 
\begin{enumerate}
\item each list has a different vertical position on the interface, 
\item each list is governed by a topic, tag, or category, and 
\item the lists generally present their items from left to right. 
\end{enumerate}

\begin{figure}[!ht]
    \centering
    \includegraphics[width=\textwidth]{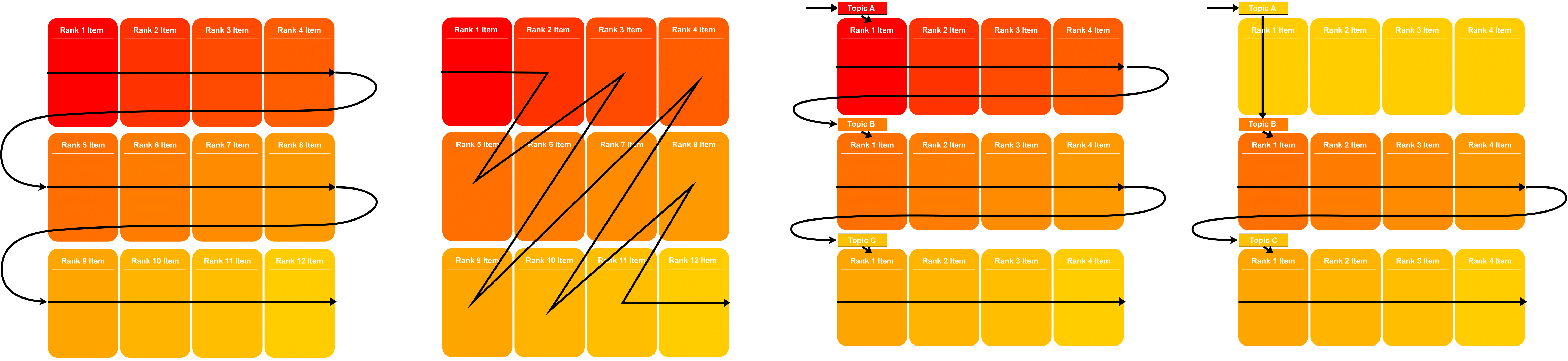}
    \caption{Examples of 2D ranked list with the standard assumed browsing behavior (far left), 2D ranked list with golden triangle browsing behavior observed in eye tracking studies (second from the left), and carousels with a topic first standard browsing behaviour (two images on the right). Color refers to item and topic relevance from most relevant (red) to least relevant (yellow).}
    \label{fig:big2}
\end{figure}

The first and second property distinguish carousels from a 2D ranked list by having a different topic for each row (in practise the rows in the carousel are spatially separated and the topic is displayed as a header for a row) and each row being a separate list. A 2D ranked list is a 1D ranked list that has been cut and stacked generally with a ranking pattern as seen in \textbf{Figure \ref{fig:big2}}. Most carousels also have an added element of complexity in that they allow display of more items than the initial presented, called swipeable carousels \cite{bendada_carousel_2020}. This allows for even more diverse browsing behaviors and the possibility of generating within-topic online recommendations based on current page feedback. 

While there is already some research on carousels, it is currently hampered by two main barriers: 1) lack of public datasets and 2) lack of empirical studies of users' browsing behavior in carousel-based interfaces.

In practise, the main problem in recommender systems is what items to present to a user when interacting with the system, i.e., how to determine the item relevance for a user (in a specific context). It was the availability of many public recommender datasets containing large amounts of user feedback \cite{harper_movielens_2016,bennett2007netflix} that propelled the research and discovery of a multitude of techniques of determining items relevance for users in a single ranked list or of learning to rank, since all these methods use click or rating data for training and evaluation. For such advancements to also happen in carousel interfaces, multiple publicly available datasets of user feedback are needed.

Another key piece of answering how to present items and determine relevancy is how users interact or browse within the recommendation system interface. The assumption that is typically used in the single ranked list is that users browse from top of the list to the bottom of the list. This top-down browsing behavior and position bias can be seen in click data \cite{joachims_accurately_2005}, while eye tracking reinforces this with empirical evidence of how a user interacts directly with the interface. Single ranked lists have the advantage of being directly linked to the presentation of web search results in information retrieval, leveraging eye tracking studies performed within the web search domain \cite{granka_eye-tracking_2004, cutrell_what_2007,pan_determinants_2004} along with eye tracking studies within the recommender system domain \cite{castagnos_eye-tracking_2010, zhao_gaze_2016, li_towards_2017, gaspar_analysis_2018}. Eye tracking studies provide a clear picture of how users interact with a system, what assumptions are reasonable to make, what biases are present, and how each should be accounted for when designing feedback models. 

For example, eye tracking studies at the time motivated the popular cascade click model \cite{craswell_experimental_2008}, which assumes that a user browses a list using a top-down approach and that items after the click are not observed and items before the click were seen (or skipped). However, determining true item examination before a click is not possible with only click data. To build upon a cascade click model, the probability of termination, the user exiting possibly due to unsatisfactory results seen so far, can also be included \cite{rahdari_ranked_2022}. 

Both item examination and termination are elements that can feasibly be learned from eye tracking data (a possible direction for future work) rather than inferring them probabilistically as is done in click models, but it is difficult to say how well eye tracking data can be generalized across users, groups, and tasks. It may not be necessary to address every possible sequence of interactions that may be learned from eye tracking data, leading to an overly complicated model, but it is imperative to confirm even the most basic assumptions and discover if there is a prominent interaction or sequence of interactions that may be beneficial to model. 
Eye tracking user studies were done in both search engine results pages and in ranked list recommendation systems and paved the way for the extensive work that has improved how user feedback is modeled. Therefore, eye tracking studies are necessary stepping stones for the development of user feedback modeling in any interface whether old or new, such as the carousel. 

While a general browsing behavior can be theorized in the case of a single 1D or 2D ranked list, it becomes much more complicated in the case of carousels, since a user can interact with topics in many ways. For example, \textbf{Figure \ref{fig:big2} } gives two possible browsing behaviors for a carousel interface: the first which is similar to that of a 2D ranked list and the second in which a user finds Topic A to be undesired and does not examine the first carousel going directly to Topic B and its carousel. This is a topic first browsing behavior, which was used in the first carousel click model \cite{rahdari_ranked_2022}. 

However, in reality browsing paths can be more complicated in the case of two dimensions. Kammerer and Gerjets \cite{kammerer_how_2010} compared user browsing behavior between a 1D ranked list and 3x3 2D ranked list across 80 subjects. For the 1D ranked list, they found that users tended to linearly browse from top to bottom, which was consistent with eye tracking studies of search engine results pages \cite{cutrell_what_2007,granka_eye-tracking_2004,joachims_accurately_2005,salmeron_comprehension_2010}. Their results in the 2D ranked list showed users browsed in a non-linear fashion, rather than going row by row or column by column, the two were mixed. 
Zhao et al. \cite{zhao_gaze_2016} found similar results in eye tracking 17 subjects during task-based exploration of 2D ranked lists, where they confirmed the F-pattern gaze hypothesis, which is also known as the "golden triangle" \cite{felicioni_measuring_2021, chierichetti_optimizing_2011} shown in \textbf{Figure \ref{fig:big2}}. 

However, there are no eye tracking studies to date for carousel interfaces only user studies \cite{jannach_exploring_2021,starke_serving_2021}. For this reason, the research project's initial contribution will be the first eye tracking study within a carousel movie recommendation setting not only to examine user behavior, but to use the insights learned for the two main contributions of designing better carousel click models and formulating learning to rank from click data, while also providing another dataset for evaluation.

\section{Research Questions and Proposed Approach}
This PhD research project seeks to address challenges in carousel recommendation and further advance the areas of carousel click models and learning to rank in carousels. Specifically, we aim to address the following identified open research questions: 

\begin{itemize}
    \item[\textbf{RQ 1:}] What are the common user browsing behaviors when presented with recommendations in a carousel and how do genre preferences impact this browsing behavior? 

    \item[\textbf{RQ 2:}] How effective is a row, column position based model for modeling clicks of users, especially as compared to the first carousel click model?

    \item[\textbf{RQ 3:}] How can we formulate and solve the problem of learning to rank directly from clicks in the carousel setting?
\end{itemize}

The first step in this project is to collect more empirical evidence on how users interact with carousel interfaces. Therefore, we will conduct the first eye tracking user study of movie carousels examining browsing behavior and the impact of genre preference on browsing (RQ 1). We plan to make this dataset publicly available, being -- to the best of our knowledge -- the first public eye tracking dataset within a (carousel) recommender system setting and the third public dataset of carousel recommenders \cite{bendada_carousel_2020, jannach_exploring_2021}. We hope that making the dataset public will help advance the field, while also helping to inform the next steps of this research project and providing more data for model validation. 

After this first step of an eye tracking user study, the two main contributions of the project will involve addressing areas in the research where carousels are lagging behind ranked lists. The first being the advancement of carousel click models, in particular the formulation of a row, column position based carousel click model (RQ 2), and the second being learning to rank directly from clicks from a carousel interface (RQ 3). 

We reference (to the best of our knowledge) all prior and current works on carousel interfaces and recommendations as related to each research question and propose our approach on how to address them.

\subsection{RQ~1: Users' Browsing Behavior with Carousels}


With regards to the research specific to recommender systems with carousel interfaces, there is much less work compared to non-carousel and web search settings. This is in part due to the presence of only a few public datasets \cite{bendada_carousel_2020, jannach_exploring_2021}: 1) the first being a dataset of n=974,960 anonymized Deezer, an online music streaming platform, user embeddings and n=862 playlist embeddings in a simulation framework with a "ground truth" display-to-stream probability and no feedback data and 2) the second a small user study of n=776 clicks in single
ranked lists and carousels. 

Additionally no eye tracking user study has been conducted with a carousel interface. As a consequence, most carousel studies have used synthetic data from Movielens and Netflix datasets\cite{harper_movielens_2016,bennett2007netflix} or created simulations of how a user may interact under a carousel setting. Two user studies (without eye tracking) have been conducted comparing ranked lists and carousel interfaces \cite{jannach_exploring_2021,starke_serving_2021}. The small amount of user studies and lack of any eye tracking study leave a gap in the literature and understanding of how users interact with the carousel interface. Eye tracking particularly can provide general browsing behaviors, item examination, and topic examination that can help motivate click models and greatly help in learning to rank, especially in learning the propensity of observations for debiasing clicks \cite{joachims_unbiased_2017}. 

To address the lack of datasets and lack of user studies in the carousel setting, the initial contribution of the doctoral research project will be an eye tracking user study to determine the browsing behavior in movie carousel interfaces and examine the impact of carousel topic preference on browsing behavior. This will be done by designing a desktop interface, similar to Netflix, initially presenting 4 different movie genres (of the total 8) carousels each with 6 displayed movie posters and the ability to swipe each carousel to display more movies and also scroll the page downwards to reveal the 4 other genres. 

30 screens are planned to be shown to at least 60 participants with the task to browse for a movie that the user would like to watch ultimately generating n=1800 clicks of feedback. 

Participants will initially be asked their preferred genres which will be used to generate 15 screens containing at least one preferred genre carousel in the initial screen (no page scrolling required) and the other 15 screens will not contain any preferred genre carousels in the initial screen. 

The purpose of this experiment is to present a naturalistic movie selection setting similar to streaming services like Netflix to better understand general browsing behavior and compare browsing behaviour between the half splits of screens to determine the impact of genre preference on browsing behaviour. We hope to examine consistent patterns in browsing behavior across the 30 screens and additional browsing behaviors that are related to genre preference. For example, the study may elucidate how users interact with the topics (genres) of the carousel, a key difference between the carousel and 2D ranked lists that may break the golden triangle browsing behavior and confirm the assumed topic browsing behavior of the carousel click model \cite{rahdari_ranked_2022}.

\subsection{RQ~2: Carousel Click Models}
Click models are generative models that seek to explain how a user interacts with a recommender system, generally some form of a ranked list. Clicks originally gathered from search engine results pages were used in the field of information retrieval to improve search engines and were naturally extended to recommender systems to do the same equivalent task, improvement of the recommendation model. The click still remains as the most common feedback (implicit or not) for both information retrieval and recommender systems, but in the field of recommender systems explicit feedback may also be gathered, such as item ranking, like/dislike, item add-to-cart, and item purchase. 

Early researchers in information retrieval believed that item relevance led to clicks and applied this to click through data gathered from search engine results pages to improve search engine results. However, in the early 2000s Joachims \cite{joachims_optimizing_2002} showed clicks were dependent on position and further research argued that position may be more important than item relevance \cite{keane_are_2008}. 

In order to better understand how and why users were clicking on search results, eye tracking studies were conducted showing the importance of position, confirming the top-down browsing behavior \cite{guan_eye_2007,cutrell_what_2007,granka_eye-tracking_2004}, and showing the correspondence between clicks and explicit judgements \cite{joachims_accurately_2005,joachims_evaluating_2007,zong_estimating_2014}. This was naturally extended to ranked lists in recommender systems and over time eye tracking studies were conducted to determine browsing behavior in 2D single ranked lists \cite{zhao_gaze_2016, kammerer_how_2010}, examine user traits \cite{chen_eye-tracking-based_2022, millecamp_classifeye_2021}, and predict gaze or interest \cite{li_towards_2017,zhao_gaze_2016}. Eye tracking studies have allowed researchers to better understand the process leading to a click and design better click models. The most popular click model is the previously mentioned cascade click model for 1D single ranked lists \cite{craswell_experimental_2008} that uses this observed top-down browsing behavior to account for position bias. 

Clicks models can allow researchers to test their assumptions on user browsing behavior, examine and discover biases, and improve ranking policies online or offline. This is why click modeling in new interfaces like the carousel is so important for the advancement of the field. For example, researchers at Deezer modeled carousel personalization as a multi-armed bandit problem with multiple plays and took inspiration from the cascade click model to deal with the problem of unobserved songs in the playlist and also integrated semi-personalization through user clustering \cite{bendada_carousel_2020}. The two most recent works in carousels by Rahdari et al. simulated user browsing behavior in a carousel interface \cite{rahdari_simulation-based_2022} and designed the first carousel click model \cite{rahdari_ranked_2022} comparing them to ranked lists to reach an analytical understanding of the prominence of carousels. There is a large gap between the single carousel click model and the many in the ranked list setting, including the  popular cascade click model and more \cite{agichtein_learning_2006, chapelle_dynamic_2009,craswell_experimental_2008,guo_click_2009,guo_efficient_2009, richardson_predicting_2007}.

In terms of click models, we wish to expand on the work of Rahdari et al \cite{rahdari_ranked_2022} and design a position based click model where a user would examine an item (i,j) at row position i and column position j with probability related to the probability of examining a row i and probability of examining a column j. This is one of many click models that can be created for the carousel interface, especially with the countless models from ranked lists that may be transitioned to carousels. The data gathered from the user study will be used to evaluate the designed click models and inform assumptions and biases taken into account when designing the models. It may also motivate novel formulations of click models based on the eye tracked browsing behaviors.

\subsection{RQ~3: Learning to Rank in Carousel Setting}
While click models are concerned with generating click data that is similar to real-life users, one of their primary goals is to help inform the process of ranking items in a list, also known as learning to rank. Beginning with a web search engine or recommendation system, from gathered clicks we would like to improve our engine or system. 

The difficulty is that it is not clear why users click a certain item or link. Assumptions can be tested by user studies or click modeling, but even then it can still be difficult to decipher a click, especially when in the context of being presented with other items. 

One approach is to ignore the context of choice of one item amongst others and just look at the feedback as positive, which is commonly done in collaborative filtering transforming click data to explicit feedback \cite{ferrari_dacrema_troubling_2021,liang_variational_2018}, usually binary consumption (1 signifies the item was clicked or rated and 0 signifies the item was unclicked and unrated). This especially makes sense in the domain of movie recommendation where it is common to have explicit feedback and users may have seen a movie outside of the streaming service or recommender system, where the choice context would be impossible to determine. Moreover, there are large databases of movie ratings that allow the recommendation of movies based on item-item, user-user, and item-user similarities. A ranked list in this case can simply be created by listing the top n most similar movies to the user in question. 

Gathering more click or rating data will help improve collaborative filtering recommendations, but recommendations may be improved further by taking into account the context of choice in ranked list presentation. Moreover, collaborative filtering models have been shown to have strong popularity biases \cite{canamares_should_2018}, which can be made worse by not taking into account ranked list position bias and other presentation biases. This creates a feedback loop of placing the most popular unseen items at the top of the ranked list, which increase their chances of being seen and then clicked, and upon retraining the collaborative filtering model will recommend these popular items even more. While this provides a straightforward approach in effectively recommending the most popular items, it makes recommendation of novel items particularly difficult.

Rather than focus on a collaborative filtering approach to ranking we consider the context of choice and learn to rank directly from the click data. The conventional problem remains the same as its earliest formulations in information retrieval \cite{joachims_optimizing_2002} by defining a risk function that aggregates the loss of a ranking of documents given a query over the query distribution, with the objective to find a ranking function that minimizes this risk. 

Learning this ranking function is commonly done through Empirical Risk Minimization \cite{joachims_optimizing_2002, joachims_unbiased_2017}. However, a problem arises in that we do not know the relevances of documents given a query, which is known as partial information learning to rank \cite{joachims_unbiased_2017}. We only have access to user feedback, which is representative of a user's relevancy judgment specific to the context when the feedback was gathered. In the case of implicit feedback (clicks), we must take into account presentation bias. Joachims \cite{joachims_unbiased_2017} addresses this by defining the \textit{propensity} of the observation, the marginal probability of observing the relevancy signal of documents given a query and the ranking presented to the user. Using a counterfactual model, an unbiased estimate of the loss of a ranking of documents given a query, presented ranking, and observed relevances can be calculated via inverse propensity scoring.  In other words, it provides a general framework for learning to rank from biased user feedback based on user's relevancy signal, the observation/examination pattern, and  propensities of observations. 

When taking into account clicks and positional bias, a position based click propensity model is used to determine the propensities of the observed clicks or the examination probabilities of each result, which can be learned by a swap-intervention experiment. This leads to an unbiased model that can learn from clicked results without assuming: 1) that unclicked results are irrelevant and 2) knowing whether the unclicked results were examined. 

Learning to rank was also formulated in recommender systems \cite{freno_practical_2017} and extended to non-contextual and contextual bandit methods. Bandit methods are simply another approach, motivated by reinforcement learning, to the problem of learning to rank balancing exploration and exploitation of recommendations \cite{mcinerney_explore_2018}. In carousels, researchers experimented on synthetic and online datasets to optimize banner carousels (carousels showing one item at a time and switching to the next after a certain time period) with contextual bandit algorithms \cite{ermis_learning_2020}. While this is the main work in learning to rank in carousels, there has been tangential work.  Felicioni and Ferrari Dacrema et al. have proposed offline evaluation protocols for carousel interfaces taking into account complementary lists \cite{felicioni_measuring_2021,felicioni_methodology_2021, ferrari_dacrema_offline_2022}. Lo et al. worked on optimizing the personalization of related item carousels present on a specific product page \cite{lo_page-level_2021}. However, as Rahdari et al. \cite{rahdari_ranked_2022} mention there is gap in the literature in terms of learning to rank directly from clicks. 

Our final contribution will be examining and working on the problem of non-bandit learning to rank with click data, as there are no approaches in the literature as of yet. This presents the challenges of having system defined queries (topics) and a governing page/session query that can affect the relevancy signals observed along with a much more complex observation pattern across multiple lists. Moreover, a risk equation would need to be formulated over the whole space of this problem with a well defined loss to allow for learning.

\section{Preliminary Results}
Preliminary work was done on using eye tracked area of interest (movie poster) dwell time as an additional source of implicit feedback to improve movie recommendations from a collaborative filtering model \cite{de_leon-martinez_eye_2023}. The PhD research project has transitioned from this particular focus of directly using eye tracking signal as feedback to improve recommender systems to using eye tracking to learn general browsing behaviors and improve click models and learning to rank due to the problem of generalizability of user, group, and task specific eye tracking signals. 

Moreover, generalizing models that learn to rank using eye tracking signals to the standard use case of recommender systems without eye tracking data poses a difficult problem that we believe needs to be addressed with large amounts of eye tracking data within a recommender system. While we seek to provide some of the data necessary, we believe that the community overall would need to focus on this problem to reach the data necessary. While this topic is no longer the focus of the project, it may be pursued secondarily especially if new approaches are discovered. 


\section{Conclusion} 
This paper presents a doctoral research project that seeks to understand user behavior in recommendation systems with carousel-based interfaces to implement new click models and develop the formulation of learning to rank from carousel clicks. The expected results of this project are to help address the lack of carousel feedback datasets, encourage the research and publication of recommender datasets with eye tracking data, and tackle the undeveloped areas of click models and learning to rank in carousel recommenders, which are lagging behind ranked list recommenders.

As the first step towards achieving the expected contributions and results, we will conduct an eye tracking user study of a carousel-based interface, the methodology of which we introduced in more detail in the paper. Any feedback during the Doctoral Symposium on the proposed ideas, study design, and models designs would be greatly appreciated.

\begin{acks}
This work was supported by Eyes4ICU, a project funded by the European Union's Horizon Europe research and innovation funding programme under grant agreement No. \href{https://doi.org/10.3030/101072410}{101072410}. I would like to acknowledge the guidance and support of both of my supervisors: Maria Bielikova and Robert Moro, and in addition Branislav Kveton for motivating the direction of this work. 
\end{acks}

\bibliographystyle{ACM-Reference-Format}
\bibliography{references}

\end{document}